\documentstyle[12pt]{article}

\newcommand{\D} {\Delta}
\newcommand{\bq} {\bar{q}}
\newcommand{\be}{\begin{equation}}
\newcommand{\ee}{\end{equation}}
\newcommand{\ba}{\begin{eqnarray}}
\newcommand{\ea}{\end{eqnarray}}
\begin{document}
\def\pct#1{(see Fig. #1.)}

\begin{titlepage}
\hbox{\hskip 12cm ROM2F-96/34  \hfil}
\hbox{\hskip 12cm hep-th/9606169 \hfil}
\hbox{\hskip 12cm \today }
\vskip .8cm
\begin{center}  {\Large  \bf  Chiral \ Asymmetry \ in \ Four-Dimensional \\
\vskip 24pt Open-String \ Vacua }
 
\vspace{1.2cm}
 
{\large \large C. Angelantonj}

\vspace{.3cm}

{\sl Dipartimento di Fisica \\  Universit{\`a} dell' Aquila \\  I.N.F.N. -
L.N.G.S.
\\ 67010 \ Coppito (AQ) \ \ ITALY}

\vspace{.7cm}

{\large \large M. Bianchi, G. Pradisi, A. Sagnotti and Ya.S. 
Stanev\footnote{I.N.F.N.  Fellow, on Leave from Institute for Nuclear Research
and Nuclear Energy, \\ Bulgarian Academy of Sciences, BG-1784 Sofia, BULGARIA.}}

\vspace{.3cm}

{\sl Dipartimento di Fisica \\ Universit{\`a} di Roma \ ``Tor Vergata'' \\ 
I.N.F.N.\ -
\ Sezione di Roma \ ``Tor Vergata'' \\ Via della Ricerca Scientifica , 1 \\ 00133
\ Roma \ \ ITALY}
\end{center}
\vskip .6cm

\abstract{Starting from the type IIB string on the $Z$ orbifold, we construct
some chiral open-string vacua with $N=1$ supersymmetry in four dimensions. The
Chan-Paton group depends on the (quantized) NS-NS antisymmetric tensor. The
largest choice,
$SO(8)
\times SU(12) \times U(1)$, has an anomalous $U(1)$ factor whose gauge boson 
acquires a mass of the order of the string scale. The corresponding open-string
spectrum comprises only Neumann strings and includes three families of chiral 
multiplets in the $({\bf 8},{\bf 12^*}) + ({\bf 1},{\bf 66})$ representation.  A
comparison is drawn with a heterotic vacuum with non-standard  embedding, and
some properties of the low-energy effective field theory are  discussed.} 
\vfill \end{titlepage}
\makeatletter
\@addtoreset{equation}{section}
\makeatother
\renewcommand{\theequation}{\thesection.\arabic{equation}}
\addtolength{\baselineskip}{0.3\baselineskip} 
\section{Introduction}

Calabi-Yau (CY) compactifications of the type IIB  string yield $N=2$
supergravity in four dimensions with $h^{(1,1)}+1$ hypermultiplets and
$h^{(1,2)}$ vector multiplets, and are naturally related via a world-sheet 
orbifold procedure 
\cite{cargese} to four-dimensional open-string vacua with $N=1$ supersymmetry. At
rational points of their moduli spaces, many CY compactifications have
microscopic descriptions as Gepner models, tensor products of $N=2$
superconformal models
\cite{gepner}.  Both these models and $N=1$ superconformal models of free fermions
\cite{fermi} are often related to orbifolds of tori \cite{dhvw} by continuous
deformations. Open descendants of rational models have been studied rather 
extensively in six dimensions \cite{bs,gp,new}, and have displayed a variety  of
new interesting phenomena \cite{tensor}. The simple rational setting of
\cite{bs}, however, is not convenient to explore the four-dimensional models,
since typical rational lattices involve (quantized) background values of the
NS-NS antisymmetric tensor that reduce \cite{torus} the size of the Chan-Paton
(CP) group.  Therefore, in this letter we return to the original setting of
\cite{ps}, essentially equivalent to \cite{gp}, and construct chiral
four-dimensional models related to the type IIB string on the $Z$ orbifold. For
notational simplicity, in all amplitudes we omit measure factors and
contributions inert under the $GSO$ projections.

\section{The $Z$ Orbifold and the Type I Superstring}

In order to construct the open descendants of the type IIB superstring on the
$Z$ orbifold, let us briefly recall the building blocks of the original
construction of \cite{dhvw}. One starts from the compactification on a six-torus
consisting of three orthogonal copies of a two-dimensional hexagonal lattice with
scalar product
\be   (e_a,e_b) = G_{ab} =  {R^2\over 3} \left( \matrix{ 2 & 1 \cr 1 & 2
\cr}\right) \quad .
\label{z3metric}
\ee  The corresponding type IIB partition function is
\be T = {| V_8 - S_8 |}^2  \ {\left( \ {{\sum \  q^{{{\alpha^\prime} \over 4}
p_{La} G^{ab} p_{Lb}}\ {\bar{q}}^{{{\alpha^\prime} \over 4} p_{Ra} G^{ab} p_{Rb}}}
\over { \eta^2(q) \eta^2( \bar{q} ) }} \ \right)}^3 \quad ,
\label{torus2b}
\ee where 
$V_8$ and $S_8$ are level-one $SO(8)$ characters, $(p_a)_{L,R} = m_a \pm {1
\over 2}G_{ab}n^b$ and $q=e^{2\pi i\tau}$. As anticipated, we begin with a
vanishing  NS-NS tensor $B$ in order to obtain a CP group of maximum size.

The $Z$ orbifold construction uses the natural $Z_3$ action on the three complex
internal coordinates $X^i \sim \omega X^i$, where $\omega = e^{{2 i \pi} \over
3}$,  that results in a total of 27 fixed points. In particular, the projection
of the untwisted sector may be expressed in terms of
\be H_{0,{\epsilon}}(q)  =  q^{1\over {12}}
\prod_{n=1}^{\infty}(1-\omega^{\epsilon} q^n)  (1 - {\bar \omega}^{\epsilon} q^n) 
\quad,
\label{untwisted}
\ee  with  $\epsilon = {0,\pm 1}$, while the twisted sectors may be expressed in
terms of
\be  H_{+,{\epsilon}}(q) = H_{-,{- \epsilon}}(q) = 
3^{-{1\over2}}{q^{-{1\over{36}}}}
\prod_{n=0}^{\infty}(1 - \omega^{\epsilon} q^{n+{1\over 3}}) (1 -  {\bar \omega}^
{\epsilon} q^{n+{2\over3}}) \quad .
\label{twisted}
\ee

Maintaining $N=2$ space-time supersymmetry in the type IIB  superstring requires
that   twists on internal bosonic coordinates, $X^i$, be properly correlated with
twists
 on internal world-sheet fermions, $\psi^i$ \cite{dhvw}. The level-one $SO(8)$
characters $V_8$ and $S_8$ associated to the transverse world-sheet fermions are
 thus to be decomposed with respect to $SO(2) \times SU(3)
\times U(1)$. To this end, it is convenient to introduce the level-one $SU(3)$
characters
$\{\chi_0,\chi_{+},\chi_{-} \}$, of conformal weights $\{0,1/3,1/3 \}$, and to 
define the supersymmetric characters $\{ A_0 , A_+ ,A_-\}$, of conformal weights
$\{1/2,1/6,1/6 \}$. The ten-dimensional $GSO$ projection then translates into
\be  V_8 - S_8 = A_0 \chi_0 + A_+ \chi_- + A_- \chi_+  \quad .
\label{gso}
\ee
$A_0$, $A_+$ and $A_-$ may be expressed in terms of the four level-one $SO(2)$
characters and of the 12 characters 
$\xi_m$ 
$(m=-5, \ldots ,6)$ of the $N=2$ superconformal model with $c=1$  (equivalent to
the rational torus at radius $R= \sqrt{12}$), of conformal weight 
$h_m={{m^2} \over {24}}$, as 
\ba A_0 &=& V_2 \xi_0 + O_2 \xi_6 - S_2 \xi_{-3} - C_2 \xi_3
\nonumber \\  A_{+} &=& V_2 \xi_4 + O_2 \xi_{-2} - S_2 \xi_1 - C_2 \xi_{-5}
\nonumber \\  A_{-} &=& V_2 \xi_{-4} + O_2 \xi_2 - S_2 \xi_5 - C_2 \xi_{-1} \quad
.
\label{As}
\ea  The spectrum of the ``parent'' type IIB string on the  $Z$ orbifold can thus
be extracted from the torus amplitude
\ba   T &=& {1 \over 3} \ \Xi_{0,0}(q) \ \Xi_{0,0}(\bq) \sum q^{{{\alpha^\prime}
\over 4} p_{La} G^{ab} p_{Lb}} {\bar{q}}^{{{\alpha^\prime} \over 4} p_{Ra} G^{ab}
p_{R b}}
 + \ {1 \over 3} \ \sum_{\epsilon = \pm 1} \ \Xi_{0,\epsilon}(q) \
\Xi_{0,\epsilon}(\bq)
\nonumber \\  & & \qquad + \ {1 \over 3} \ \sum_{\eta = \pm 1} \ \sum_{\epsilon =
0,\pm 1} \ 
\Xi_{\eta,\epsilon}(q) \ \Xi_{-\eta,-\epsilon}(\bq) \quad ,
\label{tz3}
\ea  where 
\ba
\Xi_{0,\epsilon}(q) &=& \left( {{ A_0 \chi_0 + \omega^{\epsilon} A_+ \chi_- + 
{\bar\omega}^{\epsilon} A_- \chi_+ } \over {{H_{0,\epsilon}}^3}} \right) (q) 
\nonumber \\
\Xi_{+,\epsilon}(q) &=& \left( {{ A_0 \chi_+ + 
\omega^{\epsilon} A_+ \chi_0 +  {\bar\omega}^{\epsilon} A_- \chi_- } \over
{{H_{+,\epsilon}}^3}} \right) (q) 
\nonumber \\
\Xi_{-,\epsilon}(q) &=& \left( {{ A_0 \chi_- + \omega^{\epsilon} A_- 
\chi_0 +  {\bar\omega}^{\epsilon} A_+ \chi_+} \over {{H_{-,\epsilon}}^3}}
\right)(q) 
\quad .
\label{chars}
\ea  The massless untwisted states comprise the $N=2$ supergravity multiplet, the
universal dilaton hypermultiplet and 9 additional hypermultiplets.  The twisted
sectors account for 27 hypermultiplets, one for each fixed point.  Thus, the
massless spectrum corresponds to a CY compactification with Hodge numbers
$h^{(1,1)}= 36$ and $h^{(1,2)}=0$. Two of the four scalars in each
hypermultiplet, however, arise from the R-R sector. Since their emission vertices
vanish at zero momentum, it seems more appropriate to describe them in terms of
complex antisymmetric tensors
$B^{(I)}_{\mu\nu}$, with $I=0,1,..,h^{(1,1)}$ \cite{fersab}. 

In constructing the open descendants, one starts by halving the torus amplitude
of eq. (\ref{tz3}). Since the $Z_3$ action of the target space twist is L-R
symmetric, the Klein-bottle amplitude is
\be  K = {1 \over 6} \  \Xi_{0,0}(q^2) \ \sum \  q^{{{\alpha^\prime} \over 2} m_a
G^{ab} m_b} \ + \ {1 \over 6} \ \Xi_{0,+}(q^2) \ + 
\ {1 \over 6} \ \Xi_{0,-}(q^2)  \quad ,
\label{kz3}
\ee  where $q=\exp({-2\pi \tau_2})$,  with $\tau_2$ the closed-string ``proper
time''.  Eq. (\ref{kz3}) contains only  the conventional sum over the momentum
lattice since, for generic values of $R$, the condition
${p_L}^{\omega}=p_R$ does not have any non-trivial solutions. One may thus
anticipate that the open sector includes only Neumann charges associated with the
ubiquitous nine-branes.  This should be contrasted with the $Z_2$ case
\cite{ps,gp}, where additional contributions to $K$ signal the appearance of $D$
charges in the open spectrum. The massless states in the projected closed-string
spectrum thus comprise the $N =1$ supergravity  multiplet, a universal linear
multiplet and 9 additional linear multiplets from the untwisted sector, as well
as 27  linear multiplets from the twisted sectors.

The twisted sector of the world-sheet orbifold, to be identified with the
open-string spectrum, starts with the annulus amplitude  
\ba   A &=& {{({\cal N}+{\cal M}+\bar{\cal M})^2} \over 6 } \ 
\Xi_{0,0}(\sqrt{q}) \
\sum \  q^{{{\alpha^\prime} \over 2} m_a G^{ab} m_b} \nonumber \\    & &+
{{({\cal N} +
\omega {\cal M} + \bar \omega \bar{\cal M} )^2}
\over 6} \ \Xi_{0,+}(\sqrt{q}) + {{({\cal N} + \bar\omega {\cal M} + 
\omega \bar {\cal M} )^2} \over 6} \ \Xi_{0,-}(\sqrt{q}) \quad ,
\label{az3}
\ea   where ${\cal N},{\cal M},\bar{\cal M}$ are CP multiplicities. The M\"obius
amplitude presents some subtleties connected with the proper definition of a set
of real ``hatted'' characters
\cite{bs}. Since both the twisted sectors and the  projections of the untwisted
sector are independent of the moduli $R_i$, it proves convenient to exploit the
enhanced $SU(3)^3$ symmetry at the rational point
$R_i=\sqrt{3}$ to define
\ba
\hat\Xi_{0,\epsilon} &=& \left( \hat{A}_0 \hat\chi_0 + 
\omega^{\epsilon} \hat{A}_+\hat\chi_- +   {\bar\omega}^{\epsilon} \hat{A}_-
\hat\chi_+ \right)
\left( \hat\chi_{0}^{3} - \omega\hat\chi_{\epsilon}^{3} - \bar\omega
\hat\chi_{-\epsilon}^{3}\right) \nonumber \\
\hat\Xi_{\epsilon,0} &=& \left( \hat{A}_0 \hat\chi_{\epsilon} + 
\hat{A}_{\epsilon}\hat\chi_0 -  
\hat{A}_{-\epsilon}\hat\chi_{-\epsilon}  \right) \left(
\hat\chi_{-\epsilon}\hat\chi _{0}^{2} + 
\hat\chi_{\epsilon} \hat\chi_{-\epsilon}^{2} - \hat \chi _0 \hat
\chi_{\epsilon}^{2}
\right) \quad ,
\label{hatted}
\ea  with $\epsilon = \pm 1$. This choice of signs defines a flip operator for
open strings that ensures the compatibility of direct and transverse M\"obius
channels, related by the transformation
$P = T^{1/2} S T^2 S T^{1/2}$, that simply interchanges
${\hat{\Xi}}_{0,\epsilon}$ and ${\hat{\Xi}}_{-\epsilon,0}$.   One may then verify
that
\ba  M &=& - {{({\cal N}+{\cal M}+\bar{\cal M} ) } \over 6} \  {\hat
\Xi}_{0,0} ( -
\sqrt{q} )
\ \sum
\  q^{{{\alpha^\prime}
\over 2} m_a G^{ab} m_b}
\nonumber \\  &-& {{({\cal N}+\bar\omega {\cal M} +\omega\bar {\cal M} )} 
\over 6} \ {\hat \Xi}_{0,+} ( - \sqrt{q} ) - 
 {{({\cal N} +\omega {\cal M} + \bar\omega \bar {\cal M} )} 
\over 6} \ {\hat \Xi}_{0,-} ( - \sqrt{q} )
\label{mz3}
\ea completes the open sector of the spectrum, while the three vacuum-channel
amplitudes $\tilde{K}$, $\tilde{A}$ and
$\tilde{M}$ are compatible with factorization. Demanding that unphysical massless
states decouple from the transverse channel leads to the two tadpole conditions
\ba   {\cal N}+{\cal M}+\bar{\cal M} &=& 32
\nonumber \\  {\cal N} - {1 \over 2} ({\cal M} +\bar{\cal M} ) &=& -4  \quad ,
\label{tadpolez3}
\ea   related to untwisted and twisted massless exchanges  respectively. Together
with (\ref{mz3}), eqs. (\ref{tadpolez3})  yield a CP gauge group
$SO(8) \times SU (12) \times U(1)$.  Eqs. (\ref{az3}) and (\ref{mz3}) and the
definitions in (\ref{chars}) then imply that the open sector contains three
generations of chiral multiplets in the representations
$({\bf 8},{\bf 12^*_{-1}})$ and $({\bf 1},{\bf 66_2})$. Tadpole cancellation
guarantees that this chiral spectrum is anomaly free, aside from the
$U(1)$ factor.

\section{The Anomalous $U(1)$}

As pointed out in \cite{dsw}, the $U(1)$ anomaly translates into a Higgs-like
mechanism that gives the $U(1)$ gauge field a mass of the order of the string
scale
$M_s$.  Denoting by $tr$ the trace in the fundamental representation, the
anomaly  six-form is
\be A_6 \sim  F \left( {1 \over 12} F^2 + tr(F_{12}^2) - tr(F_8^2) \right) \quad ,
\label{anomaly6}
\ee where $F$, $F_{12}$ and $F_8$ are field strengths of $U(1)$, $SU(12)$ and
$SO(8)$. The resulting consistent anomaly is canceled by a  Green-Schwarz
mechanism induced by
$L_{GS}= B\wedge F$ and by the Chern-Simons coupling
\be   H = dB + {1 \over 12} A \wedge F + \omega_{3(12)} - \omega_{3(8)} \quad .
\ee   It is convenient to perform a duality transformation, turning $B_{\mu\nu}$ 
into a pseudo-scalar axion $\beta$, and combine it with the dilaton $\phi$ into a
complex scalar $S=\beta+ie^{-2\phi}$.  In the heterotic string, at tree level $S$
describes a non-linear $\sigma$-model on the coset $SL(2,R)/U(1)$  with K\"ahler
potential $K = - \log(S - \bar S)$ and enters the gauge kinetic function $f_{ab}
= S
\delta_{ab}$ \cite{witred}.  Gauging the holomorphic isometry generated by 
$\partial/\partial\beta$ leads to the coupling
$\partial_{\mu}\beta \rightarrow D_{\mu}\beta=\partial_{\mu}\beta + k A_{\mu}$,
and  the gauge variation of $\beta$, $\delta \beta = - k \alpha$, qualifies it as
a St\"uckelberg field. The supersymmetric completion of these couplings is simply
$K = - \log(S - \bar S +  i k V)$.

In compactifications of the $SO(32)$ heterotic string with standard embedding,
the  (generically) anomalous spectrum of $U(1)$ charges is  responsible for the
generation of a Fayet-Iliopoulos $D$-term at one-loop. Massless scalars charged
under $U(1)$ that are singlets of the unbroken gauge group
$SO(26)\times SU(3)$ allow one to shift the vacuum with broken supersymmetry to a
nearby supersymmetric vacuum, giving masses to $U(1)$ charged particles as well
to  the $U(1)$ gauge field
\cite{dsw}. Other would-be axions cannot acquire couplings  similar to
$\beta$ because of non-renormalization theorems \cite{dsw}.

Although it is  reasonable to expect that anomaly related terms in the effective
lagrangian  turn into one another under heterotic - type I duality \cite{tse}, 
the simple strong - weak coupling duality $\phi_I = -\phi_H$ in $d=10$ 
\cite{vw}  (that would already affect the loop-counting arguments of \cite{dsw})
changes deeply in lower dimensions.  Indeed, since the dilaton belongs to the
universal sector of the  compactification, the relation between the heterotic and
type I dilaton in $d$ dimensions is determined by dimensional reduction to be
\be
\phi^{(d)}_I \ = \ {6-d \over 4} \ \phi^{(d)}_H \ - 
\ {{(d-2)} \over {16}} \log \det G^{(10-d)}_H \quad ,
\ee   where $G^{(10-d)}_H$ is the internal metric in the heterotic-string frame,
and there is a crucial sign change at $d=6$.  The independence of $\phi_H$ and
$\phi_I$ in $d=6$ is not a surprise \cite{six}. From  the rational construction
of \cite{bs}, as well as from the orbifold approach in
\cite{gp,new}, it is well known that type I models exist with different numbers
of tensor multiplets, a setting with no analogue in  perturbative heterotic
compactifications on $K3$.  Moreover, open descendants of type IIB Gepner models
lead naturally to low-energy spectra  without tensor multiplets
\cite{ournew}. In these cases the dilaton belongs to a  hypermultiplet to be
identified, on the heterotic side, with one of the moduli of the $K3$
compactification.  In four dimensional
$N=1$ models, the dilaton lies in a linear multiplet on both sides, and heterotic
- type I duality appears to be related to chiral - linear duality.  The presence
of an anomalous $U(1)$ suggests that R-R fields, that flow in the transverse
channel and can thus take part in a generalized Green-Schwarz mechanism
\cite{tensor}, correspond to charged scalars on the heterotic side. 

\section{A Candidate Heterotic dual}
 
In order to substantiate these arguments, we now construct a heterotic model with
(almost) the same perturbative spectrum as the open-string model of the previous
section.

This heterotic model corresponds to compactification on the $Z$ orbifold with
non-standard embedding of the spin connection into the $SO(32)$ gauge group. The
twist needed to achieve the desired result consists of four copies of the basic
twist
$(1/3,1/3,1/3)$ \cite{dhvw} and clearly satisfies the level matching constraints
of modular invariance. As a result, $SO(32)$ is broken to
$SO(8)\times U(12)$, the CP group of the type I model. Moreover, the untwisted
charged spectrum coincides with the open-string spectrum of the type I model. 
Indeed, the twist splits the heterotic R-moving fermions, $\tilde\lambda^I$, into
three sets. Those inert under the twist,
$\tilde\lambda_a$, with $a=1,\ldots,8$, those that pick a  phase $\omega$,
$\tilde\lambda_r$, and those that pick a phase $\bar \omega$,
$\tilde\lambda_{\bar r}=\delta_{\bar rs}
\tilde\lambda^s$, with  $r,s=1,\ldots,12$. The emission  vertices of the  charged
massless scalars in the $({\bf 8,12^*})$  and $({\bf 1,66})$ are
$V^{ir}{}_a = \psi_i \tilde\lambda^{r} \tilde\lambda_a$ and $V^i_{[rs]} = \psi_i
\tilde\lambda_{r} \tilde\lambda_s$, respectively. 

A striking feature of the heterotic model is that twisted massless scalars are
charged  with respect to the gauge group. To show this, it proves very convenient
to  exploit the conformal embedding of $U(12)$ into $SO(24)$ at level one induced
by the branching
${\bf 24} = {\bf 12_1} + {\bf 12^*_{-1}}$. Denoting by $X_n$ the characters of
the integrable representations based on the $n$-fold antisymmetric products of
the fundamental representation of $SU(12)$ and by $\xi_m$ the $N=2$ superconformal
characters defined before eq. (\ref{As}), one finds the decomposition
\be O_{32} + S_{32} = O_8  O_{24}+V_8 V_{24} + C_8 C_{24} + S_8 S_{24} = B_0 +
B_+ + B_- \quad ,
\label{os32}
\ee where 
\ba  B_0 \!\!\!\! &=& \!\!\!\! O_8(X_0\xi_0+X_6\xi_6)+C_8(X_0 \xi_6+X_6\xi_0)  
+V_8(X_3\xi_{-3}+X_9\xi_3)+S_8(X_9\xi_{-3}+X_3\xi_3)
\label{so24su12} \\  B_+ \!\!\!\! &=& \!\!\!\!
O_8(X_2\xi_{-2}+X_{8}\xi_4)+C_8(X_8\xi_{-2}+X_2\xi_4)
+V_8(X_{11}\xi_1+X_5\xi_{-5})+S_8(X_{11}\xi_{-5}+X_5\xi_{1})
\nonumber \\  B_- \!\!\!\! &=& \!\!\!\!
O_8(X_{10}\xi_2+X_{4}\xi_{-4})+C_8(X_4\xi_2+X_{10}\xi_{-4}) +V_8(X_1\xi_{-1}+X_7
\xi_5)+S_8(X_1\xi_5+X_7\xi_{-1}) \ .
\nonumber
\ea   Embedding the $Z_3$ twist in the $Z_{12}$ center of $SU(12)$ assigns a phase
$\omega^{\epsilon}$ to $X_{3k+\epsilon}$, with $\epsilon=0,\pm 1$. An S modular
transformation  on the resulting projection
$B_0+\omega B_+ +\bar\omega B_-$ yields the R-moving twisted sector.  Combining
it with the L-moving sector and taking into account the Casimir energy $\D
E=1/3$, one obtains massless particles only from $O_8 X_0 \xi_{-4}$ and
$C_8 X_0 \xi_2$. The 27 blowing up modes of the fixed points, associated with $O_8
X_0$, that together with the 9 complex scalars from the untwisted sector
represent the moduli of the $Z$ orbifold, are thus charged ($m=-4$ in the chosen
units) with respect to the  anomalous $U(1)$. This is precisely what we are
suggesting for the massless R-R scalars from the unoriented closed string 
spectrum in the type I model.  The only difference between the two low-energy
spectra lies in the presence, in the heterotic case, of additional chiral
multiplets in the  $({\bf 8}_c,{\bf 1})$ representation of the surviving gauge
group, $SO(8)\times SU(12)$.

\section{Low Energy Interactions}

One can easily compute the $\beta$ functions of the
$SO(8)$ and $SU(12)$ factors, and find
$\beta^{SO(8)} = 9$, $\beta^{SU(12)}=-9$. The $SO(8)$ factor, not asymptotically
free, is not necessarily problematic in String Theory,  since one has an
effective ultraviolet cutoff, $M_s$, of the  order of the Planck mass.  The
$SU(12)$ interactions may be responsible for interesting non-perturbative effects
\cite{susyrep}. Denoting by $\Phi^{is}_{b}$ and $\chi^k_{[rs]}$ the chiral
multiplets in the
$({\bf 8,12^*})$ and $({\bf 1,66})$ representations of the CP group, the cubic
superpotential for the charged matter fields is fixed by
 gauge symmetry to be
\be  W = y \ \delta^{ab} \ \epsilon_{ijk} \ \Phi^{ir}_{a} \Phi^{js}_{b} 
\chi^k_{[rs]} \quad .
\ee  The Yukawa coupling $y$ is a constant, independent of the moduli fields of 
the  closed-string spectrum, since the matter fields belong to the untwisted
sector of
 the $Z$ orbifold \cite{dfms}.  The manifest global $SU(3)$ symmetry is expected
to be broken by higher-order terms, that we neglect for the time being. The other
renormalizable interactions are encoded in the $D$-terms. Denoting by 
$A^i$ and $F^i$ the v.e.v.'s of the scalar fields in the chiral multiplets
$\chi$  and $\Phi$ and assuming minimal kinetic terms, one gets
\be  D = \sum_i 2 {\rm tr} (A^{\dag}_iA^i ) - {\rm tr} (F^iF^{\dag}_i )
\ee   for the anomalous abelian factor,
\be   D^A = {\rm tr} (F^i T^A F^{\dag}_i ) = ({\bf T}^A )^{a}{}_{b} D^b{}_a 
\ee   for $SO(8)$, where $A=1,\ldots ,28$ and ${\bf T}^A$ are generators in the 
vector representation, and
\be   D^I = \sum_i 2 {\rm tr} (A^i {\bf t}^I A^{\dag}_i) - {\rm tr}  (F_i^{\dag} 
{\bf t}^I F^i) = ({\bf t}^I )_{r}{}^{s} D_{s}{}^{r}
\ee   for $SU(12)$, with $I=1,\ldots ,143$ and ${\bf t}^I$ the generators in the 
anti-fundamental representation.

In order to investigate further breakings of the gauge symmetry associated with 
flat directions of the scalar potential, the $D$-term conditions
$D^A=0$ and $D^I=0$ must be supplemented with the $F$-term conditions
\ba {\partial W \over \partial F^{ir}_{a}} &=& 2 y
\epsilon_{ijk}  F^{js}_{a} A^k_{[rs]}=0
\nonumber \\  {\partial W \over \partial A^k_{[rs]}} &=& y \epsilon_{ijk}
\delta^{ab} F^{ir}_{a} F^{js}_{b} = 0 \quad .
\ea   The condition $D=0$ is not to be imposed, because of the one-loop 
Fayet-Iliopoulos $D$-term.  For one family, similar problems have been studied in
the literature \cite{susyrep}.   In our model with three families, with no loss
of generality one can set
\be  F^1 = \pmatrix{f^1\cr 0}\qquad {\rm and} \qquad  A^1 = \pmatrix{a^1 & 0 \cr
0 & b^1} \quad ,
\label{fmatrix}
\ee  where $f^1$, $a^1=-(a^{1})^{T}$ and $b^1=-(b^{1})^{T}$ are (generically
invertible) complex matrices of dimension $8\times{8}$, $8\times{8}$ and
$4\times{4}$  respectively.  The $F$-term conditions then imply that all the
$F^i$ and $A^i$ are of the same form as (\ref{fmatrix}). $f^i$ and $a^i$  can
then break
$SO(8)\times SU(8)$ to
$U(1)^4$, compatibly with the  $F$ and $D$ term conditions above. On the other
hand, the three $b^i$,  that belong to the ${\bf 6}$ (complex vector)
representation of
$SU(4) \sim SO(6)$, generically break it to $SU(2)$.

Non-perturbative effects are expected to generate corrections to the tree-level
superpotential. Instanton calculus in the original $SU(12)$  is not adequate in
this case since 
$N_f=3 > {N_c\over N_c-3} = 4/3$, \cite{susyrep}.  Holomorphy, symmetry arguments
and limiting behaviors, however, are powerful tools to constrain the low-energy
effective lagrangian \cite{intseib}. In particular, the appearance of the
$SU(12)$ gauge singlet chiral multiplets $U = \chi\Phi^2$,  already present in
the tree-level superpotential,  as well as $V = \chi^{30} \Phi^{24}$, expected in
the dynamically generated superpotential, seems unavoidable.  Other composite
chiral fields, as well as real superfields, may be needed to gauge part of the
original flavor  symmetry and to achieve a consistent low-energy description
\cite{intseib}.

\section{Comments}

Let us discuss the effect of a quantized NS-NS antisymmetric tensor background,
$B_{ab}$, in the orbifold planes.  The quantization condition, related to the
requirement of left-right symmetry in the toroidal compactification, was
discussed in
\cite{torus}, where it was found that a $B_{ab}$ of rank $r$ reduces  the CP
multiplicity to $32/2^{r \over 2}$. Up to now, in order to keep the CP group as
large as possible, we have discarded this possibility.  However, it is simple to
construct a rational model based on the $Z_3$ orbifold of the $E _6$ maximal
torus \cite{masphd}, that requires a $B_{ab}$ with $r=6$. The type IIB  parent
theory is built out of 27 generalized characters, $Y_k$, with identity
$Y_0 = A_0 \chi_0^4 + A_+ \chi_-^4 + A_- \chi_+^4$, that expose a global 
$SU(3)^4$ symmetry. The diagonal and the charge-conjugation modular invariant
lead to different open descendants, but in both cases the maximal CP group is
$SO(4)$. In the charge conjugation case, the massless closed spectrum includes 36
linear  multiplets, while the open spectrum is pure super Yang-Mills.  In the
diagonal case, the massless closed spectrum includes 12 vector multiplets and 24
chiral multiplets, while the open spectrum comprises 27 chiral multiplets in the
adjoint representation of the CP group, so that the resulting theory is weakly
coupled at low-energies. Other choices of modular invariant torus amplitudes
result in different numbers of vector, linear and  chiral multiplets in the closed
spectrum, as well as in different patterns of CP symmetry breaking.  All these
models are not chiral. The intermediate possibilities for $B_{ab}$, $r=2$ and
$r=4$, correspond to non-vanishing tensor background in one or two orbifold
planes, and lead  to CP groups $SO(8)\times U(4)$ (with three generations in the
$({\bf 8},{\bf 4})$ and
$({\bf 1},{\bf 10})$) and $U(4)$ (with three generations in the ${\bf 6}$). 

Phenomenological prospects for type I vacua where not among the purposes of this
work. It is conceivable, however, that some chiral type I vacuum may accommodate
the standard model of particle physics.  The natural appearance of three
generations in the $Z_3$ case is rather amusing in this respect, and the symmetry
breaking pattern along flat directions is very rich. Thus, one may embed a
phenomenologically more appealing
$SU(5)$ in the $SU(12)$ factor of the CP group. However, it seems rather
difficult to get rid of the  remaining $SO(8)$ and to obtain a minimal spectrum
of ${\bf 10}$'s and  {\bf 5}$^*$'s.

\vskip 24pt
\begin{flushleft} {\large \bf Acknowledgments}
\end{flushleft}

It is a pleasure to thank I. Antoniadis, C. Bachas, S. Ferrara, and G.C. Rossi
for stimulating discussions. A.S. would like to acknowledge the kind hospitality
of the CERN Theory Division  while this work was being completed. This work was 
supported in part by E.E.C. grant CHRX-CT93-0340.

\end{document}